\documentstyle[prc,aps,psfig]{revtex}
\def\Journal#1#2#3#4{{#1} {\bf #2}, #3 (#4)}

\def\NPA{{Nucl. Phys.} A}
\def\NPB{{Nucl. Phys.} B}
\def\PLB{{Phys. Lett.}  B}
\def\PR{{Phys. Rep.}}
\def\PRL{Phys. Rev. Lett.}
\def\PRC{{Phys. Rev.} C}
\def\PRD{{Phys. Rev.} D}

\def\ZPC{{Z. Phys.} C}

\font\sevenrm=cmr7

\def\P{{\mathbf p}}

\def\a{\alpha}
\def\b{\beta}
\def\d{\delta}
\def\e{\epsilon}

\def\p{\pi}
\def\q{\theta}

\def\n{\nu}

\def\t{\tau}

\def\cm{{\cal M}}

\def\llra{\longleftrightarrow}

\def\del{\mbox{$\partial$}}
\def\ncdot{\!\cdot\!}
\def\srm#1{\mbox{\sevenrm #1}}

\def\eff{\srm {eff}}

\def\be{\begin{equation}}
\def\ee{\end{equation}}
\def\bea{\begin{eqnarray}}
\def\eea{\end{eqnarray}}
\def\eref#1{Eq.~(\ref{#1})}
\def\ederef#1#2{Eq.~(\ref{#1}) and (\ref{#2})}

\def\fref#1{Fig.~\ref{#1}}
\def\bfig{\begin{figure}}
\def\efig{\end{figure}}

\newcommand{\ncom}{\newcommand}
\ncom{\vo}[1]{{\mathbf #1}}
\ncom{\vmo}[1]{{|\mathbf #1|}}
\ncom{\vt}[2]{({\mathbf #1}-{\mathbf #2})}
\ncom{\lan}{\langle}
\ncom{\ran}{\rangle}
\ncom\nonum{\nonumber \\}
\ncom\fx{}
\ncom\gsim{\mbox{\raisebox{-0.6ex}{\ $\stackrel {>}{\sim}$\ }}}
\ncom\lsim{\mbox{\raisebox{-0.6ex}{\ $\stackrel {<}{\sim}$\ }}}

\ncom{\half}{{1\over 2}}
\ncom{\third}{{1\over 3}}
\ncom{\fourth}{{1\over 4}}
\ncom{\fifth}{{1\over 5}}
\ncom{\sixth}{{1\over 6}}

\ncom\Tg{T_{eq\; g}}
\ncom\Tq{T_{eq\; q}}
\ncom\qg{\q_g}
\ncom\qq{\q_q}

\begin{document}
\draft
\preprint{WU-B 97/13}

\title{$\a_s$ DEPENDENCE IN THE EQUILIBRATION\\ 
IN\\ RELATIVISTIC HEAVY ION COLLISIONS}

\author{S.M.H. Wong}

\address{Fachbereich Physik, Universit\"at Wuppertal,
D-42097 Wuppertal, Germany}

%\date{20 February, 1997}

\maketitle

\begin{abstract}

The dependence of the equilibration of the parton plasma
on the value of the strong coupling is studied in Au+Au 
collisions at LHC and at RHIC energies. With increasing
coupling, the following are found to happen:
1) both thermal and chemical equilibration speed up, 
2) in the final degree of equilibration, only
quarks and antiquarks show obvious improvements but 
not gluons and 3) the plasma cools much more rapidly.
The deconfinement phase transition will therefore takes 
place sooner and it naturally results in the shortening 
of the parton phase of the plasma. The exact duration 
of this phase is however sensitive to the value of 
the coupling. A change from $\a_s=0.3$ to $\a_s=0.5$, 
for example, reduces the lifetime of the parton phase 
at LHC by as much as 4.0 fm/c. The total generated entropy
is another sensitive quantity to the coupling. Larger 
values of $\a_s$ will lead to entropy reduction and 
therefore reduction both in the duration of the mixed
phase, assuming there is a first order deconfinement 
phase transition, as well as in the final pion 
multiplicity. It is shown that the common choice of
$\a_s=0.3$ is not a good value for the entire duration
of the evolution given that the system undergoes 
substantial changes from the beginning to the time 
that the deconfinement phase transition is about to 
take place assumed to be at $T_c \sim 200$ MeV. Instead, 
by using a more consistent simple recipe, the system is 
allowed to decide its own strength of the interactions which 
evolves with the system as it should. With this approach, 
$\a_s$ increases with time and this leads to acceleration 
in the equilibration even as equilibrium is near. This 
is opposite to the behavior of the equilibration of a 
molecular gas or ordinary many-body system where the 
interaction strength is fixed. In such system, 
the net interactions will slow down as the system is 
near equilibrium. 

\end{abstract}

%\pacs{PACS numbers: 25.75.-q, 12.38.Mh, 12.38.Bx, 24.85.+p}
\pacs{WU B 97/13 \ \ \ (to appear in Phys. Rev. C)}

\section{Introduction}
\label{sec:intro}

With the asymptotic freedom of QCD, one expects quarks and 
gluons to behave almost as free particles at very high
energies and under extreme conditions. Such extreme conditions
as believed to be found in the early universe can, to a
limited extent, be recreated in the laboratories in the
experiments of heavy ion collisions. As highly energetic and 
relativistic matter collides at 200 GeV/A at RHIC and 
6.3 TeV/A at LHC, nucleons lose their individual identities 
in favor of a gas of partons. A main goal of the experiments
is to establish beyond doubt the existence of this parton
plasma. In order to do so, distinctive signs in the 
guise of particle signatures must be looked for. Numerous
works have already been devoted to these. Also of importance 
is the temporal development of the parton plasma which 
directly influences the various particle signatures. In our 
previous investigation into the equilibration of this QCD 
plasma \cite{wong1,wong2}, we have shown that, as in agreement 
with previous works \cite{geig&mull,geig,geig&kap,biro&etal1}, 
chemical equilibration in the partonic mixture cannot be 
completed by the time that the deconfinement phase transition 
sets in. However, we also pointed out that kinetic 
equilibration might also not be as quick, or perhaps one 
should say, not as perfectly equilibrated as one would have 
liked. A thermalization time within 1.0 fm/c is unrealistic 
as shown in \cite{geig,geig&kap}. It was pointed out in 
\cite{eskola} that very fast equilibration for gluons, at 
least, based on estimate of the transverse energy deposited 
in the central collision region, might be possible depending 
on what parton distribution was used. However, one should be 
wary of the fact that initially the matter is highly 
compressed, so even though thermalization may approximately 
be achieved, the expansion may drive it out of equilibrium. 
The question then is whether this gluon early thermalization,
if it can be achieved, is a transient or a maintainable
thermalization. We have shown that in our previous work
\cite{wong1,wong2}, expansion can indeed drive out the
early thermalization which is to be recovered progressively 
only later. 

Unlike in a vacuum, in a dense QCD medium, collective 
effects will provide for infrared screening \cite{weld,klim}
and so we have no need for an arbitrary soft momentum cutoff.
This feature reduces the dependency of our investigation
on the number of external parameters. And in fact, apart
from the obvious initial inputs, the only remaining
variable which one has a certain freedom to choose is $\a_s$.
Since after all, we are doing a perturbative calculation,
a small $\a_s = 0.3$ was chosen, which corresponds to an 
average momentum transfer of $Q \sim 2.0$ GeV and 
$\Lambda_{QCD} \sim 200$ MeV. One can see from previous
works on chemical equilibration 
\cite{geig&kap,biro&etal1,shury&xion},
particle production reduces the temperature $T$ and this
lasts over several fm/c during which $T$ drops by
several hundreds of MeV. Consequently, the average parton
energy also varies considerably. As a result, we do not
and cannot expect the average momentum transfer to remain
at around $Q \sim 2.0$ GeV. Therefore $\a_s$ should also 
vary during the equilibration and evolution of the parton
plasma. This effect has not been taken into account.
If thermalization is very fast, one can perhaps argue
for a roughly constant $\a_s$ during thermalization but
certainly not during chemical equilibration when
the system changes considerably. We have plotted in
\fref{f:ave}, the average parton energies for quarks 
and gluons during the evolution of the plasma in 
our previous investigation \cite{wong2}. As can be 
seen, assuming the average momentum transfer is of 
the order of the average parton energy,
$Q \sim \e /n$, one can neither expect $Q$ to stay at around
$2.0$ GeV at LHC nor at RHIC. We therefore investigate
the dependence of our previous results on equilibration 
on $\a_s$. This we carry out in two ways. The first is
to use various fixed $\a_s$ and the second is to use an 
$\a_s$ determined by the system. The later approach means
we do not choose a value for $\a_s$ but let the system
decides what it should be. Since the system is evolving,
the resulting coupling will evolve with the system.
These will be explained in the following sections.

\bfig
\centerline{
\hbox{
{\psfig{figure=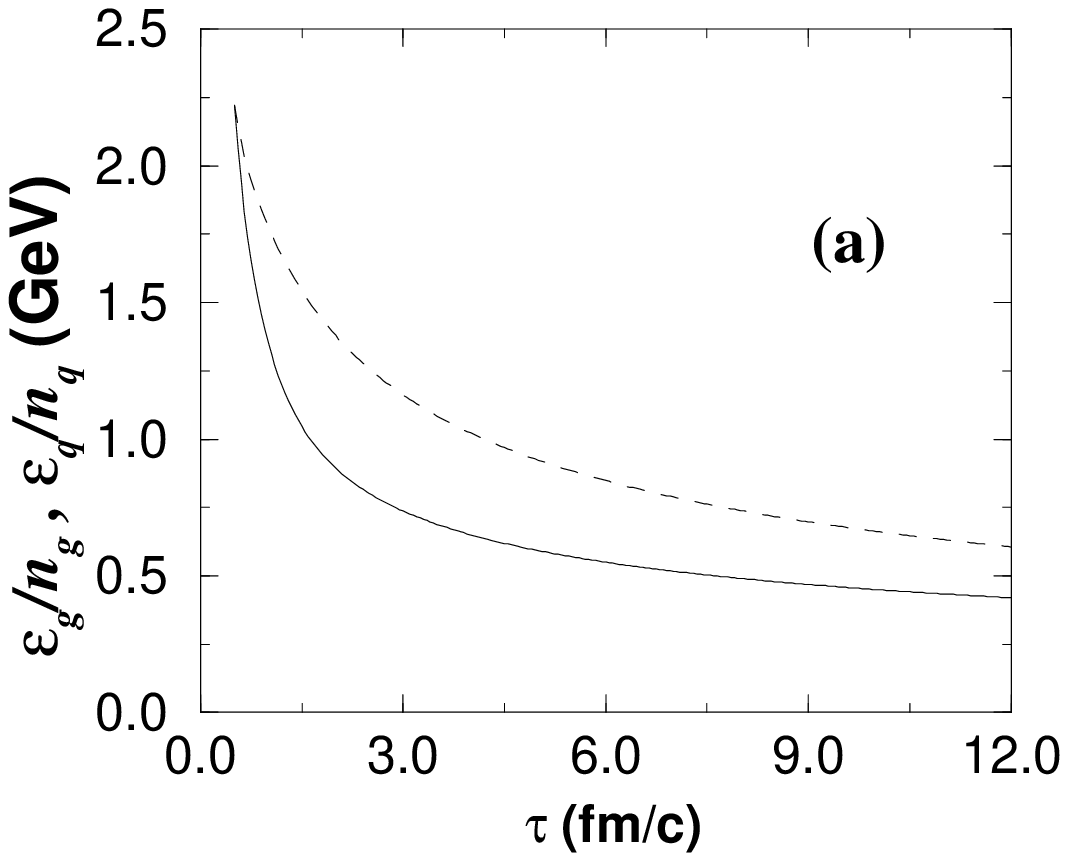,width=2.3in}} \ 
{\psfig{figure=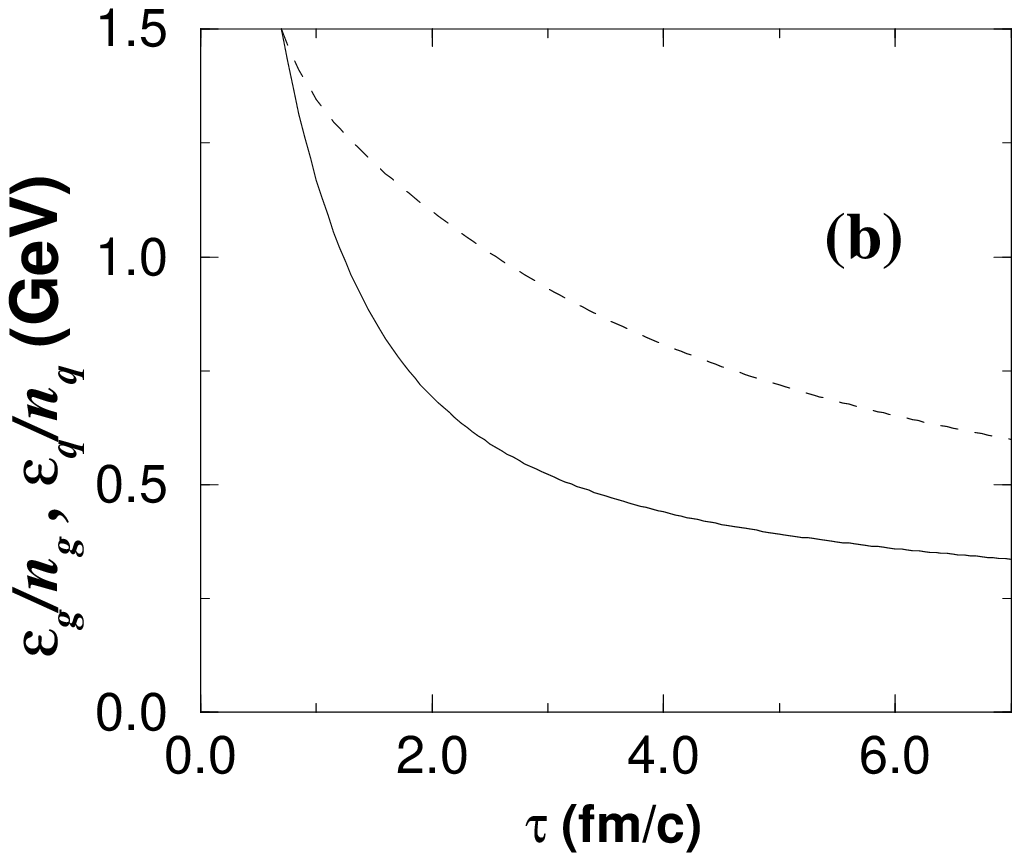,width=2.3in}}
}}
\caption{The evolution of the average parton energy of gluon
(solid line) and quark or antiquark (dashed line) at
(a) LHC and (b) RHIC with $\a_s=0.3$.}
\label{f:ave}
\efig

In Sec. \ref{sec:eq}, we recapitulate briefly our method 
and basic equations. We comment on the questions related
to the possible inclusion of a mean field term and
possible role played by instabilities. What values to
use for $\a_s$ in our investigation and how to obtain
an evolving coupling are explained. In Sec. \ref{sec:res},
results on the effect of $\a_s$ on the equilibration
will be shown and consequences discussed. We then show
that a plasma governed by QCD is no ordinary many-body
system.

\section{The evolution equations}
\label{sec:eq}

Quantum effects have proved so far to be hard to include
in all its details 
\cite{winter,heinz1,elze&etal,elze&heinz,elze,heinz2,heinz3,groot}
and interference in many particle interactions are largely
absent (except in a limited sort of way in 
\eref{eq:g_no_source} below). One can at best, at present, 
to investigate the equilibration in heavy ion collisions 
in its full form by semi-classical means.

Our basic equation is the Boltzmann equation which can, in 
Baym's manner \cite{baym}, be rewritten as 
\be {{\del f_i} \over {\del \t}} \Big |_{p_z \t}
    =C_i(p_\perp,p_z,\t)
\label{eq:baymeq}
\ee
$i=g,q,\bar q$, by the assumption of one-dimensional boost 
invariant longitudinal expansion in the very central region 
of the collision of two highly relativistic heavy ions.
The collision terms $C_i$ on the right hand side include
sums over all relevant interactions and we approximate
it by the relaxation time approximation, which is
expedient for our purpose and has been used in simpler
studies of thermalization 
\cite{heis&wang1,gavin,kaj&mat,heis&wang2}. 
Explicitly, it is written in the form
\be C_i(p_\perp,p_z,\t)=- 
    {{f_i(p_\perp,p_z,\t)-f_{eq \; i}(p_\perp,p_z,\t)} 
     \over \q_i(\t)} \; 
\label{eq:relaxapp}
\ee
where $f_{eq}$ is the full equilibrium distribution and 
is a function of the equilibrium temperature $T_{eq}$. 
Because we are considering an expanding system,
this approximation is not sufficient to close the 
equations. $T_{eq}$ and $\q$ remain functions of time.
In any case, we need input from QCD which is obtained by
explicitly constructing $C_i$ also from QCD interactions 
\cite{wong1,wong2}. We use the same set of interactions as before
\be gg \llra ggg \; \; , \; \; \; gg \llra gg \; , 
\label{eq:ggi}
\ee 
\be gg \llra q\bar q \; \; , \; \; \; g q \llra g q \; \; , 
    \; \; \; g\bar q \llra g\bar q \; ,
\label{eq:gqi}
\ee
\be q\bar q \llra q\bar q \; \; , \; \; \; qq \llra qq \; \;  , 
    \; \; \; \bar q \bar q \llra \bar q \bar q \; .
\label{eq:qqi}
\ee
The collision terms are constructed from the well known
vacuum matrix elements of the above interactions at leading 
order in $\a_s$ \cite{cut&siv} but rendered infrared safe by 
medium screening. These screening effects are put in by hand
in terms of the Debye screening and quark medium 
mass. They are calculated from the distributions $f_i$ and are
therefore functions of $\a_s$ as well as $\t$. Admittedly,
they are only part of the screening effects since they have
no momentum dependence. But for our purpose, they are sufficient
to provide the right order of magnitude for the screening. 
With these $\a_s$ dependent masses, the collision terms 
become more complicated functions of $\a_s$. The explicit
form of the infrared screened matrix elements can be found in
\cite{wong1,wong2}. Apart from screening, other medium effect
will also have to be included, that is the 
Landau-Pomeranchuk-Midgal (LPM) suppression of gluon radiations 
or absorptions due to multiple interactions. This is partially 
incorporated in the two to three gluon multiplication collision 
term in the form of a theta function 
\cite{gyul&wang,gyul&etal,baier&etal}. This collision term 
as appeared in the collision entropy rate per unit volume is 
\bea \left ({{\del s_g} \over {\del \t}} \right )_{coll}^{gg\llra ggg} 
     & = & \fx \fourth (2\p)^4 
     \n^2 \prod^5_{i=1} {{d^3 \P_i} \over {(2\p)^3 2 p^0_i}} 
     |\cm_{gg\rightarrow ggg}|^2 \d^4(p_1+p_2-p_3-p_4-p_5) 
     \ln \left ({{f_1 f_2 (1+f_3)(1+f_4)(1+f_5) } \over 
      {(1+f_1)(1+f_2) f_3 f_4 f_5 }} \right )  \nonum
     \fx & & \fx \times
    [f_1 f_2(1+f_3)(1+f_4)(1+f_5)-f_3 f_4 f_5(1+f_1)(1+f_2)] \,
     \q (\Lambda-\t_{QCD}) \; .
\label{eq:g_no_source}
\eea
where $\Lambda$ is the gluon mean free path,
$\t_{QCD}= \sqrt{s} (p_1+p_2) \ncdot p_5 
/(4 p_1 \ncdot p_5 p_2 \ncdot p_5)$ is the gluon formation 
time of the radiated gluon with momentum $p_5$
and $s =2 p_1 \ncdot p_2$ is the squared of the centre-of-momentum 
energy of the parent gluons. The gluon mean free path 
is a function of $\a_s$ as well as the Debye screening mass 
\cite{wong1,biro&etal1}. The resulting dependence on $\a_s$ 
of this is more complicated than the binary interaction terms.
Combining these explicit collision terms and that of the
relaxation time approximation, one can solve for $T_{eq \, i}$ 
and $\q_i$ at each instance in time and hence the $f_i$ 
distributions which depend on this two variables can be determined. 

As we explained briefly in our previous works 
\cite{wong1,wong2}, our $f_{eq}(\t)$ is the momentary 
``target'' equilibrium distribution at which the particle 
distribution of the system will eventually settle,
if one is able to stop the expansion at $\t$. 
This can be seen in the analytic form of the 
solution to the approximation Eq. (4) in \cite{wong2}. 
At large time compared to $\q$, the 
solution is dominated by the second term and
the integrand in this term is dominated by the
upper limit of the integral. One can approximate the
integral by evaluating the integrand at the peak
and multiply by the ``width'' of this peak, which is 
approximately given by $\q$. This means that 
the solution will tend to the equilibrium distribution
at large times. Therefore one should not confuse our approach 
with the Chapman-Enskog method of linearizing the Boltzmann 
equation. In fact, the linearization of this method
does not give a collision term of the simple form of the 
relaxation model. Also the leading particle 
distribution of the Chapman-Enskog expansion does not
have the same physical meaning as our $f_{eq}$.
In the Chapman-Enskog case, the leading distribution
is the best fitted local distribution to the system at 
any moment in line with the locally equilibrated
hydrodynamical description of the method
but ours is rather what the system would
at any time like to reach and we try to describe
what will happen early on in relativistic heavy ion
collisions and therefore before the hydrodynamic expansion
phase. The collision model by itself contains no information
about QCD, but because it is taken to model the
collision terms and so can be equated to the latter for
fixing the parameters of the distribution.

At this point, we would like to comment on the lack of a
mean field term in our basic equation \eref{eq:baymeq} which
is often a point of criticism. In \cite{mrow1,mrow2}, it
was shown that unstable collective plasma modes might
develop via chromoelectromagnetic mean fields when the
particle momentum distribution is anisotropic. 
Anisotropy will no doubt be featured in the early stage 
of heavy ion collisions which may give rise to 
instabilities. As worked out in \cite{mrow1,mrow2}, the
time scales of the instabilities are earlier than our
initial time $\t_0$ both at LHC and at RHIC. Therefore
these instabilities will influence our initial
inputs. They then become part of the many uncertainties 
associated already with the initial conditions.
Their effects on the equilibration can then be 
studied as part of the initial condition dependence. 
By the time that the evolution starts, collisions 
are important and the derivations in \cite{mrow1,mrow2} 
are no longer applicable. In any case, the mean field
term in the Vlasov-Boltzmann equation does not generate
entropy and without collisions, the mean field term 
cannot bring about equilibration. Therefore we are
doubtful that it can be very important for equilibration 
in a direct way. It may have some indirect effects as 
suggested in \cite{mrow1} but that needs further studies 
to clarify.

To study the dependence of equilibration on $\a_s$, we
evolve the plasma using other values. The previous results
were obtained with $\a_s =0.3$. In order to make the effects
prominent and unambiguous, we choose $\a_s =0.5$ and 
$\a_s=0.8$. With such large $\a_s$ and in particular 
$\a_s =0.8$, one can no longer trust leading order 
calculations, our aim is to make the influence
of $\a_s$ manifest. In any case, we are not after quantitative
but qualitative results. Apart from these values, as mentioned 
in the introduction and shown in \fref{f:ave}, the average 
parton energies vary over a rather large range during the 
evolution and so also should the momentum transfers. To
complete this study, we then use a coupling which
evolves with the system. This is done by using the 
following recipe. Since two colliding partons each carrying
the average parton energy will have a maximun momentum
transfer equals to twice the average parton energy so we
can assume the average transfer is of the order of the
average parton energy
\be Q \sim < \e_g + \e_q + \e_{\bar q} >/
    <n_g + n_q +n_{\bar q} >   \; .
\ee
Then the strong coupling is given by the one-loop running 
coupling formula $\a_s (Q) = 4 \p /\b_0 \ln (Q^2/\Lambda_{QCD}^2) $.
We choose an average value $\Lambda_{QCD} =235$ MeV \cite{pdg} 
and $n_f =2.5$. As already mentioned, this last approach
eliminates the coupling as the remaining external parameter.

\section{Results and discussions}
\label{sec:res}

We use the same initial conditions $T_0$, $l_{0 \, i}$, 
$\e_{0 \, i}$, $n_{0 \, i}$, where $i=g, q, \bar q$ as before 
\cite{wong2} to compare with our previous results.
One notes that these values from HIJING  \cite{Gyu1,Gyu2,Gyu3} 
have small initial fugacities which is partly responsible for 
the not-so-well quark chemical equilibration. One could try
multiplying the initial fugacities by a factor to compensate 
for this as done in \cite{lev&etal,wang} and also recently
in \cite{mull&etal}. Here we concentrate only on the 
effects of the variation of $\a_s$ and not worry ourselves
about the initial conditions. 

Our plots are produced with $\a_s =0.3,0.5,0.8,\a_s^v$. The
last denotes the coupling which varies with the evolution
according to the recipe given in the preceding section. 
To look at the effects of $\a_s$ on equilibration, we examine
the parton fugacities $l$, the longitudinal to transverse
pressure ratios $p_L/p_T$ and then also the temperature
estimates $T$ of the each parton component of the plasma. 
The first give us information about the parton composition,
the second reveal the state of the kinetic equilibration
of the system and the last tell us about the possible lifetime
of the parton plasma. In \fref{f:fug}, \fref{f:pres} and 
\fref{f:temp}, we plotted these results.

\bfig
\centerline{
\hbox{
{\psfig{figure=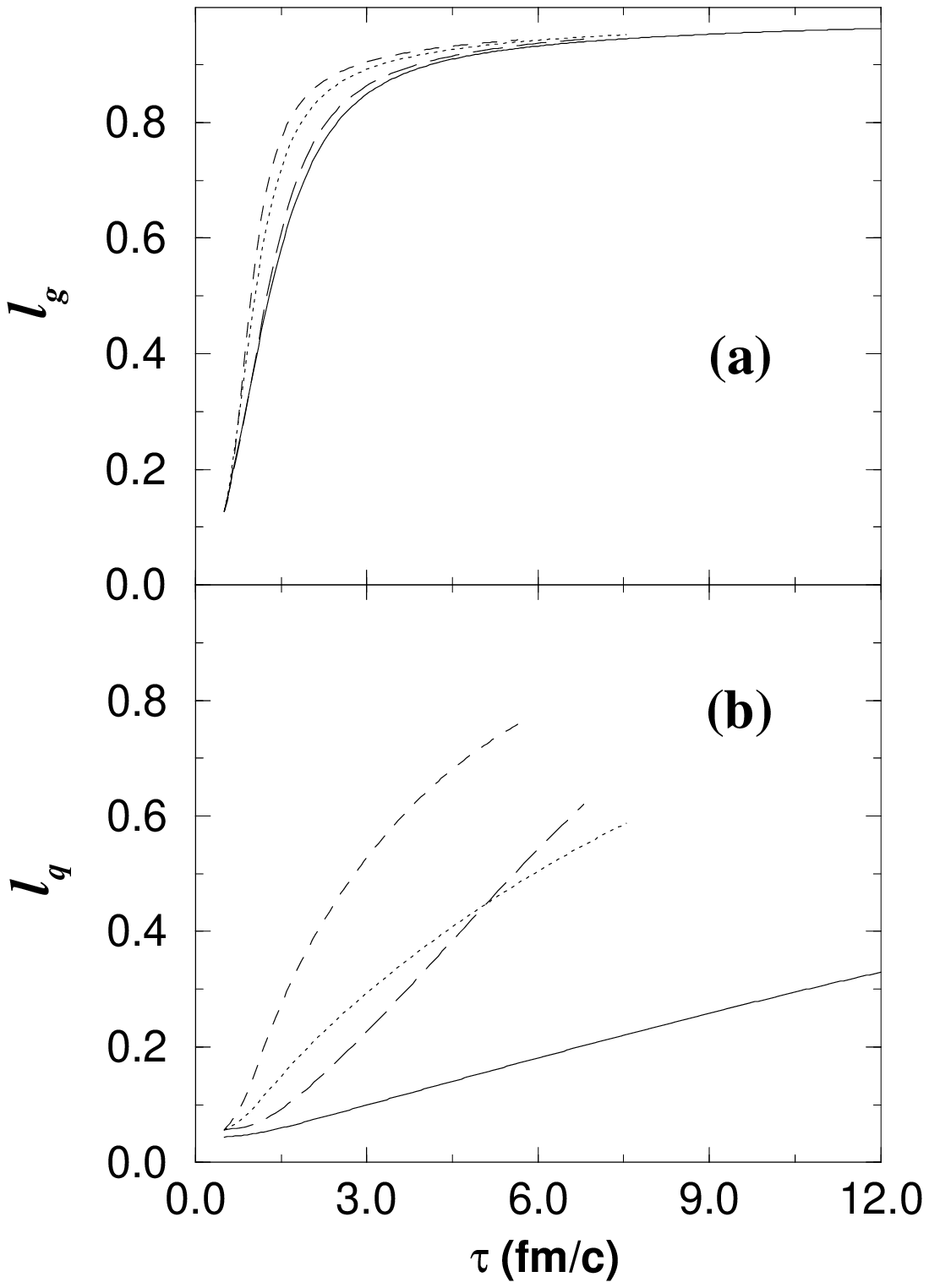,width=2.3in}} \ 
{\psfig{figure=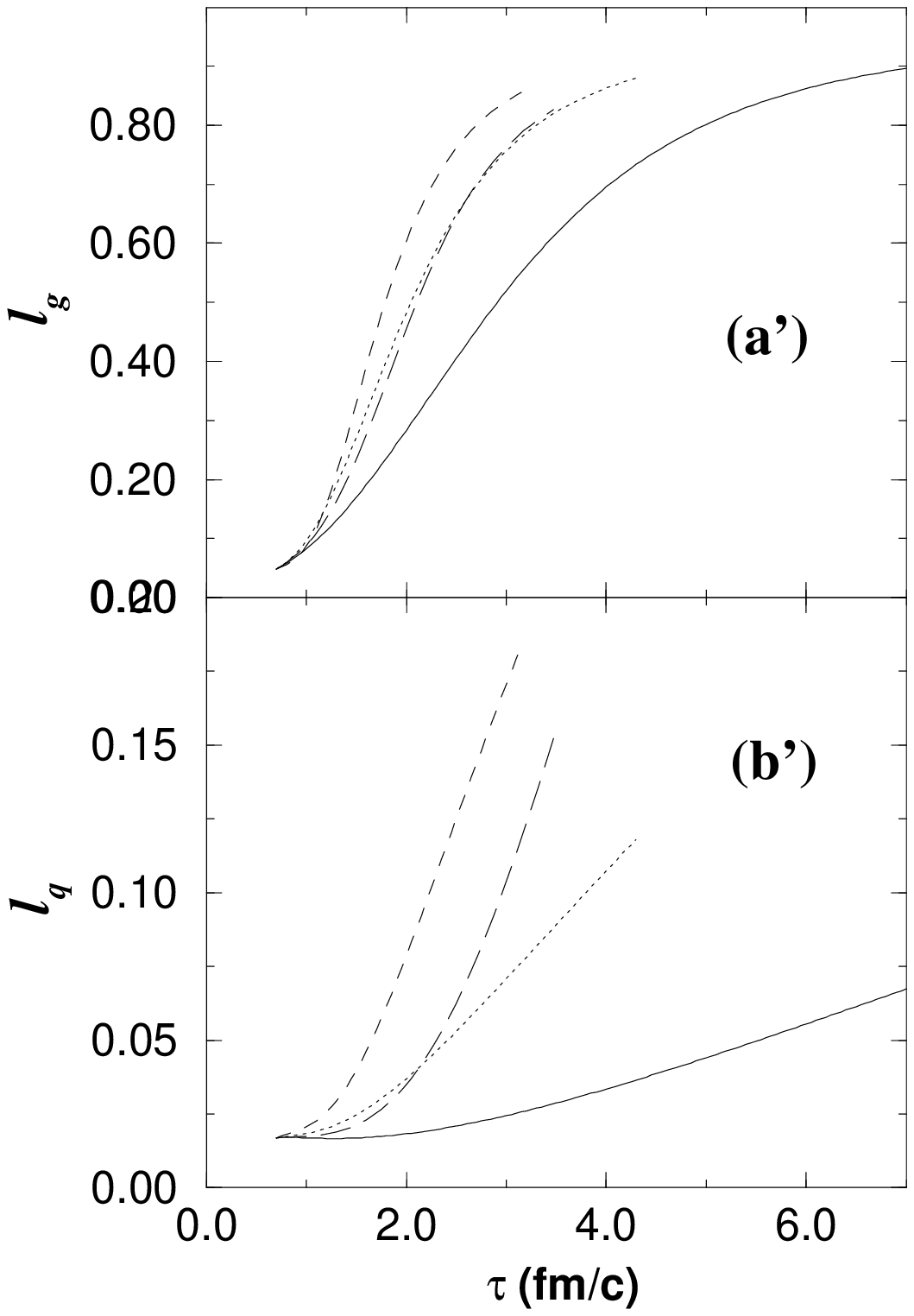,width=2.3in}}
}}
\caption{Chemical equilibration of (a) gluons and (b) quarks
with various values for the coupling: $\a_s=0.3$ (solid),
0.5 (dotted), 0.8 (dashed) and $\a_s^v$ (long dashed) at
LHC. The (a') and (b') figures are the same at RHIC.
Increasing coupling improves the quark final degree of chemical
equilibration much more than that of the gluon.}
\label{f:fug}
\efig

With varying $\a_s$, the fugacity evolution is as in 
\fref{f:fug}. The curves shift towards the upper left hand
corner with increasing $\a_s$. This is the same for gluons 
\fref{f:fug} (a) and (a') and for quarks \fref{f:fug} (b) and (b').
As can be seen, larger $\a_s$ leads to faster approach towards 
full chemical equilibration. Curves with larger $\a_s$ rise
faster. For gluons, it takes less time to achieve approximately
the same degree of chemical equilibration. Whereas for quarks,
the final fugacities are improved by 1.5-2.0 times at LHC and
a somewhat larger factor of 1.8-2.8 at RHIC. These enhancement
factors clearly depend on how close to equilibrium the
previous $\a_s =0.3$ results are. When it is farther from 1.0
as in the fermion case at RHIC, the factors are largest
and when it is close or very close as in the gluon case,
there are not much improvements. Or rather there is not much
room for improvement because it cannot go further than
full equilibration and the undoing effects of the back reactions 
are important at this stage of the equilibration. The effect 
of larger $\a_s$ is to shorten the lifetime of the
parton phase of the plasma only in this case. 

Similar situation is also found in the ratios of longitudinal
to transverse pressure which is a check of isotropy of
parton momentum distributions and a test of kinetic 
equilibration. In \fref{f:pres}, we plot the pressure ratios
$p_L/p_T$ as well as $\e/3 p_T$ both for quarks (b) and (b')
and for gluons (a) and (a') at LHC and at RHIC respectively. 
The top set of curves in each case is the $\e/3 p_T$ plots. 
In these plots, curves with larger $\a_s$ are closer to the
top in general. In other words, they are closer to full
kinetic equilibrium. As in the case of fugacities, 
kinetic equilibration is clearly faster as the amount of
time required to reach the same or a higher degree of
equilibration is shorter. However, the final situations for
gluons are about the same. The improvements for quarks are again
much clearer. Both $p_L/p_T$ and $\e/3 p_T$ plots show the
same tendency. It becomes obvious that increasing $\a_s$ 
improves the equilibration of quarks and antiquarks much 
more than that of the gluons. These improvements and faster
equilibration are however at a price. One can see that 
the curves with larger couplings are stopped earlier and that 
is because the price to be paid is more rapid cooling for 
larger $\a_s$. This can be seen more clearly in the plots 
of the estimated temperatures in \fref{f:temp}.

\bfig
\centerline{
\hbox{
{\psfig{figure=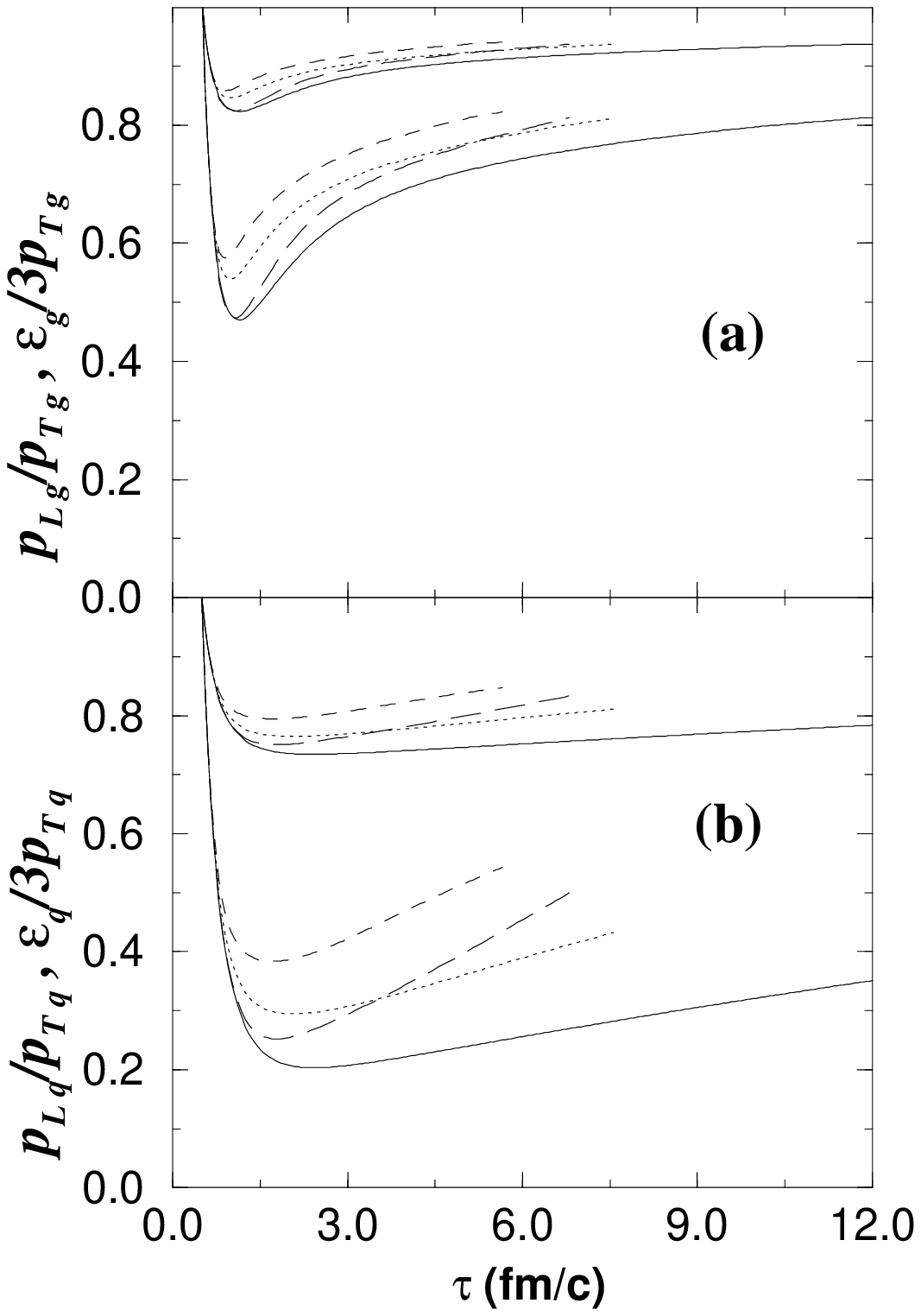,width=2.3in}} \ 
{\psfig{figure=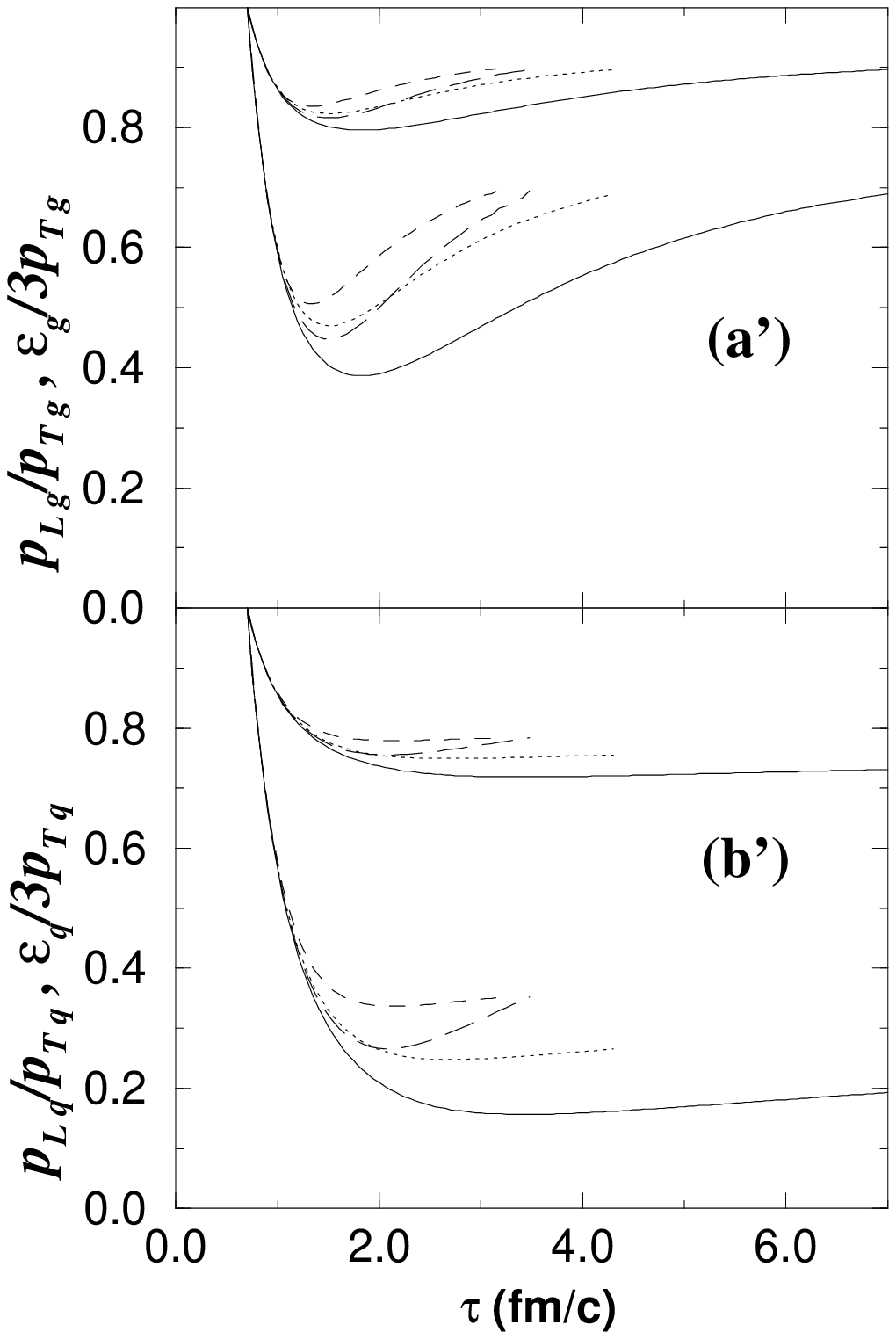,width=2.3in}}
}}
\caption{Using the ratios of the longitudinal pressure and a 
third of the energy density to the transverse pressure to 
check for isotropy in momentum distribution and therefore
kinetic equilibration. The bottom (top) set of four curves
in each figure is for the pressure (energy density) to
pressure ratio. The assignments of the coupling to the
curves are $\a_s=0.3$ (solid), 0.5 (dotted), 0.8 (dashed)
and $\a_s^v$ (long dotted). Figures (a) and (a') are
for gluons and (b) and (b') for quarks at LHC and at RHIC
respectively. Faster kinetic equilibration is seen
everywhere with larger $\a_s$ but improvement in the 
final degree of thermalization is essentially reserved 
for the fermions and not for the gluons.}
\label{f:pres}
\efig

One remarks from \fref{f:temp},
that the effect on the lifetime is considerable as a
shift from $\a_s=$0.3 to 0.5 shortens the time at which the
quark temperatures drop to 200 MeV from 12.0 fm/c to 8.4 fm/c
at LHC in \fref{f:temp} (b) and from 7.0 fm/c to 4.3 fm/c at RHIC
in \fref{f:temp} (b'). The reduction on this same duration of the 
gluon temperatures is less and is only about 2.0 fm/c at LHC 
in \fref{f:temp} (a) and 1.0 fm/c at RHIC in \fref{f:temp} (a') 
at maximum. Although gluons always cool faster than
quarks due to the combined effects of the expansion and the
loss of gluons to quarks and antiquarks, the cooling of
the fermions are, like the fugacities and pressure 
ratios, affected more by the coupling. In all, the duration
of the parton phase of the plasma is very sensitive
to the value of the coupling. To have to choose a value of
the coupling by hand is almost equivalent to choosing the 
results. So it may be more consistent by the arguments
already given to let the system determines its strength of
the interactions.

\bfig
\centerline{
\hbox{
{\psfig{figure=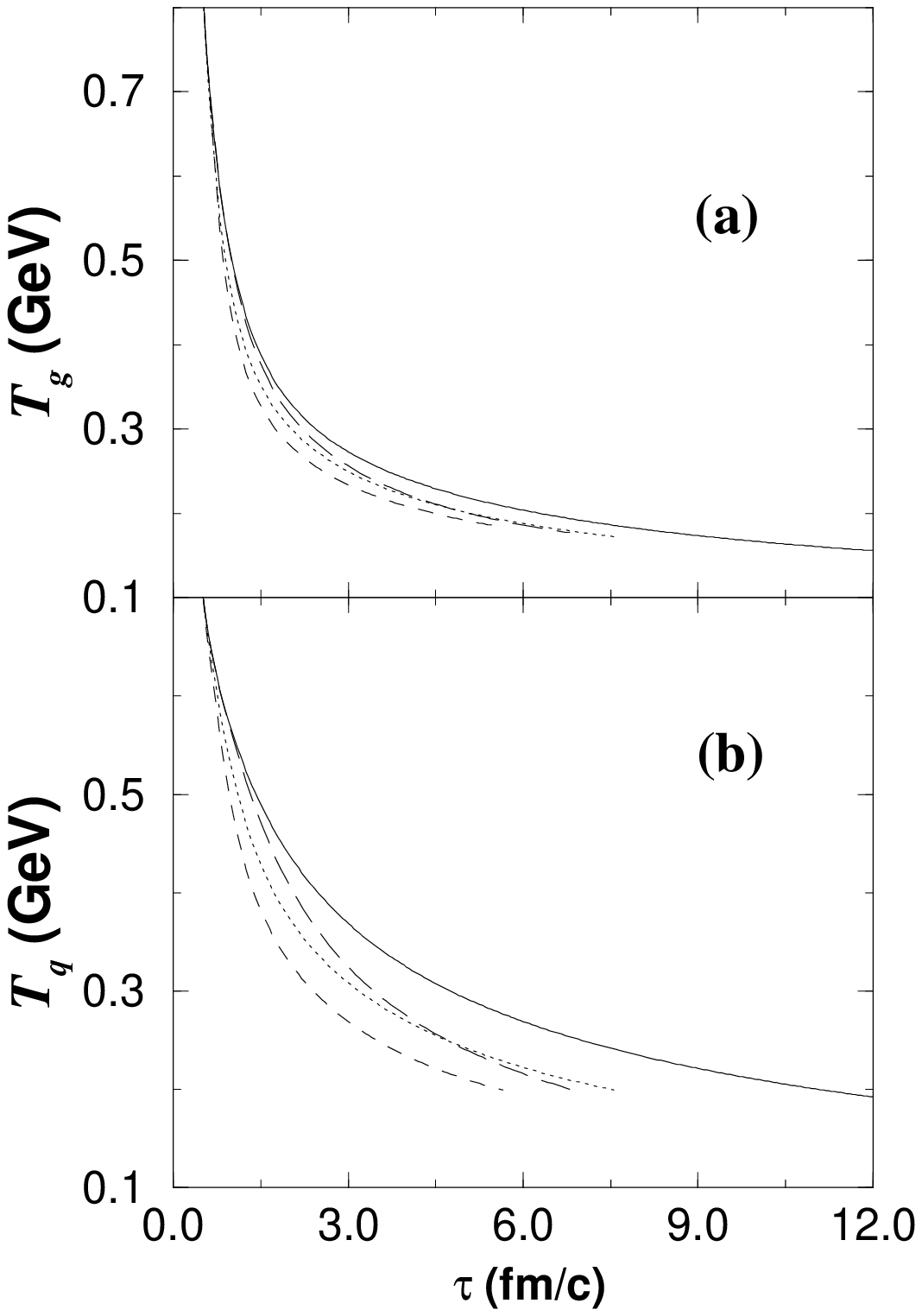,width=2.3in}} \ 
{\psfig{figure=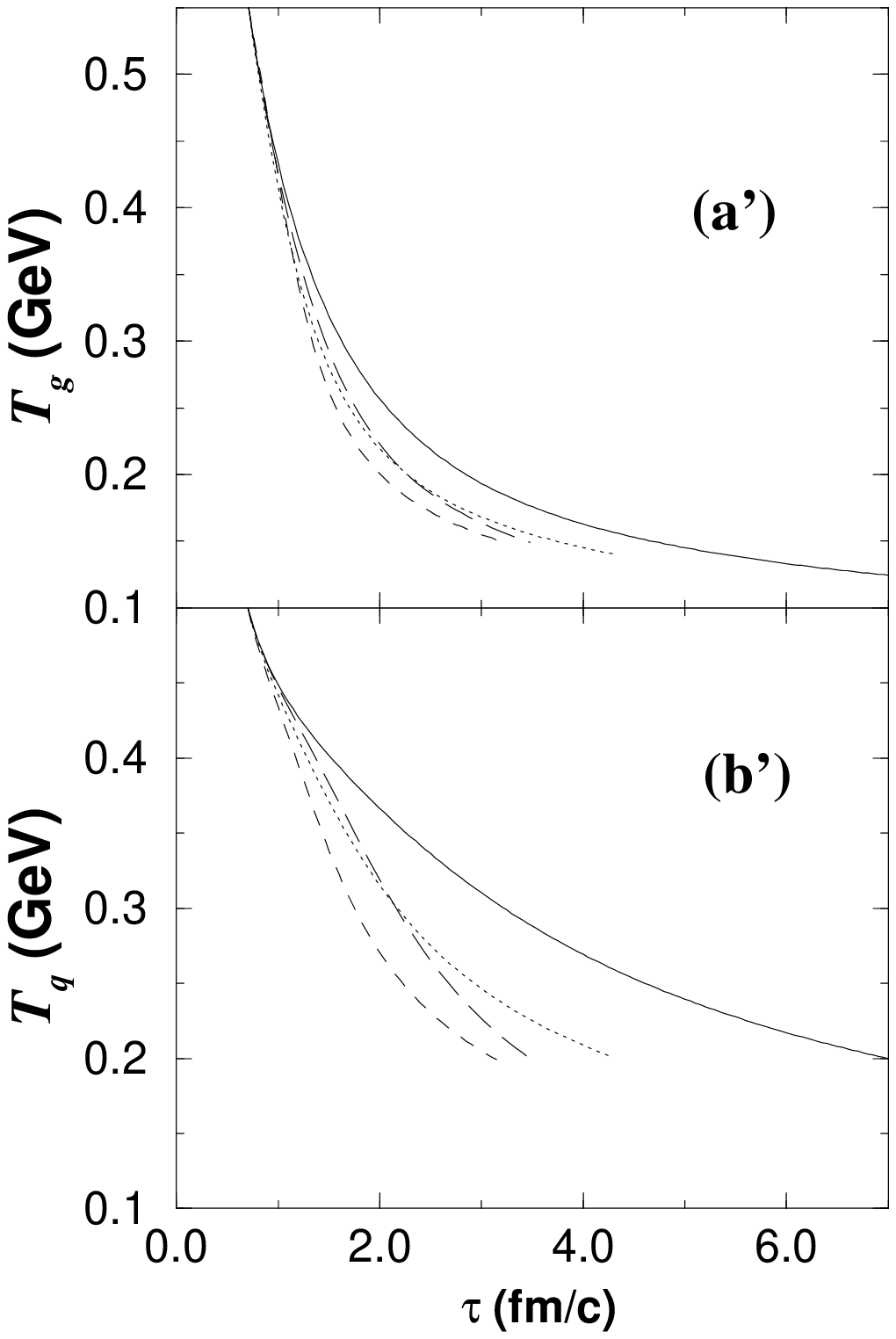,width=2.3in}}
}}
\caption{The time variations of the estimated temperatures
of (a) and (a') gluons and (b) and (b') quarks at LHC
and at RHIC respectively. These temperatures drop faster
with increasing coupling. The different values of the
coupling are assigned to the curves in the same
way as in \fref{f:fug} and \fref{f:pres}.}
\label{f:temp}
\efig

For the reasons discussed in Sec. \ref{sec:intro} and
in the previous paragraph, the case of $\a_s$ evolving with
the system, $\a_s^v$, interests us particularly.
In the plots \fref{f:fug}, \fref{f:pres} and \fref{f:temp}, 
these curves shift across the constant $\a_s$ 
``contours'' with increasing $\t$. Since the strength
of the interactions changes with the evolution, the results
are progressive departures from the previous and they
improve progressively upon those in the sense that the
fugacities, the pressure ratios are larger and 
therefore closer to equilibrium at the expense of more 
rapid cooling and shortened lifetime. Since $\a_s^v$
shifts towards larger values of $\a_s$, higher order terms
will have to be included at some stage and eventually
the problem will become non-perturbative. In this work,
we try not to worry about higher orders and just examine the
results at leading order. 

We have seen the results of how $\a_s$ affects the approach 
to equilibrium. They tell us about chemical 
and kinetic equilibration are speeded up and the final
degrees of these two aspects of equilibration have been
altered but they do not tell us much about how the partons
are behaving with increasing coupling. In this sense, 
collective variables are much more suitable for this
purpose. In any case, it would be interesting also to 
see how the collective variables are affected by the 
coupling.

In \fref{f:eprod}, the variation of the products of the
parton energy density $\e_i$ with $\t^{4/3}$ are plotted.
It would be helpful to think of each parton component
of the plasma to be subjected to an effective longitudinal
pressure $p_{L\, \eff}$, so that the equations for the
energy densities, which follow from 
\ederef{eq:baymeq}{eq:relaxapp} \cite{wong2}, become
\be  {{d \e_i} \over {d \t}} 
   + {{\e_i + p_{L\,i\, \eff}} \over {\t}}
   = 0
\ee
with the effective pressures given by
\be  p_{L\, i\, \eff} = p_{L\,i} 
   + {{\t (\e_i-\e_{eq\, i})} \over \q_i} \; .
\label{eq:press_eff}
\ee
In \fref{f:eprod} (a) and (a'), the effective pressure of
the gluons tends to increase with $\a_s$ and be larger than
one-third of the gluon energy density hence the decreasing
tendency of $\e_i \t^{4/3}$. 
The opposite is true for the quark effective pressure in 
\fref{f:eprod} (b) and (b'). In \fref{f:pres}, we have already
seen that the longitudinal pressures for all partons are less than 
a third of the corresponding energy density so the second term
in \eref{eq:press_eff} must be positive (negative) and 
increasing (decreasing) with the coupling for gluons (quarks). 
In other words, the net energy transfer from gluons to quarks
and antiquarks is positive and increasing with $\a_s$. 
This variation in the net energy transfer is accompanied by a 
corresponding increase in the loss of gluon number and 
gluon entropy and similarly an increase in the gain of 
quark-antiquark pairs and quark entropy. 
These are shown in \fref{f:nprod} and \fref{f:s}.

\bfig
\centerline{
\hbox{
{\psfig{figure=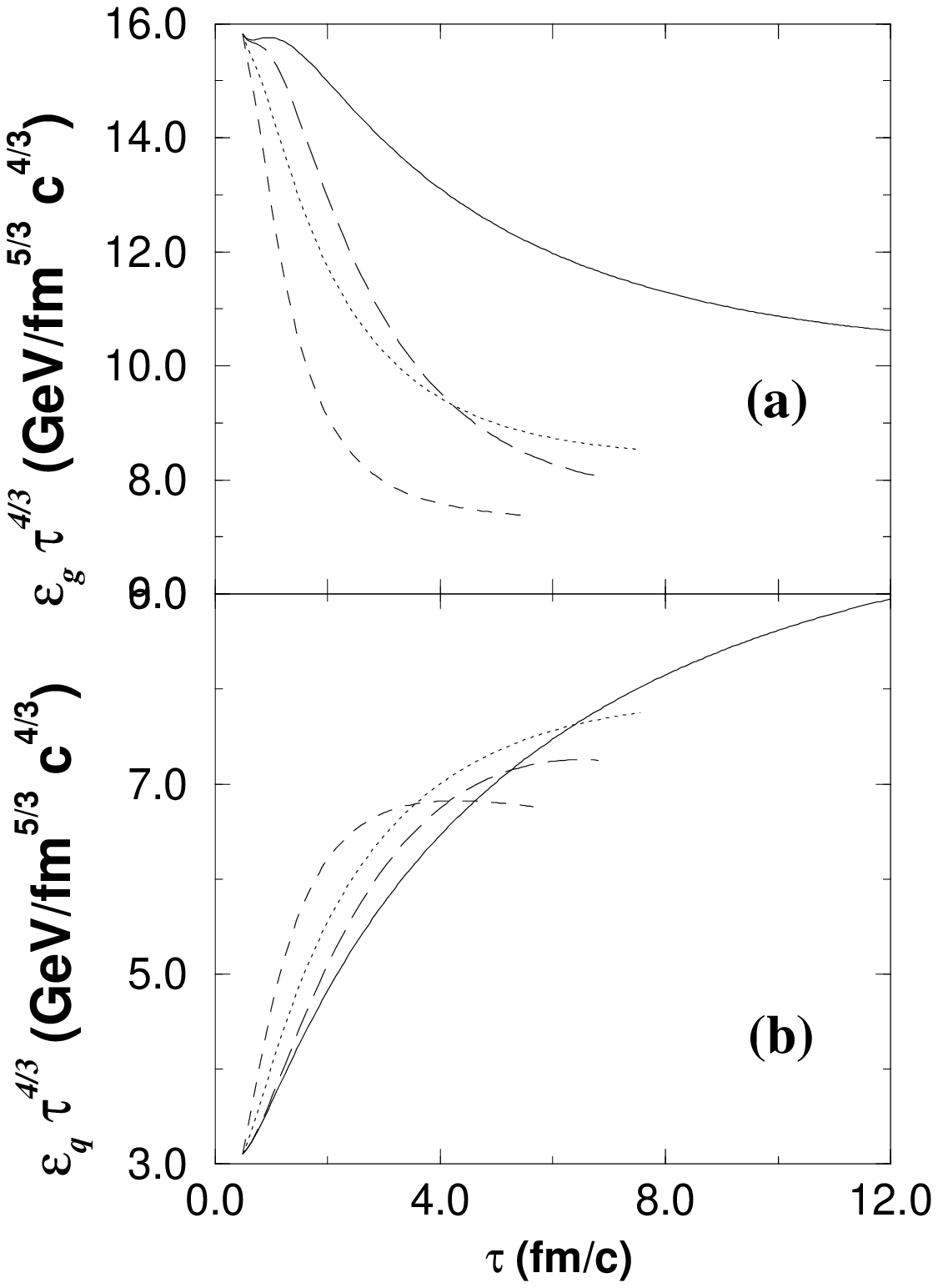,width=2.3in}} \ 
{\psfig{figure=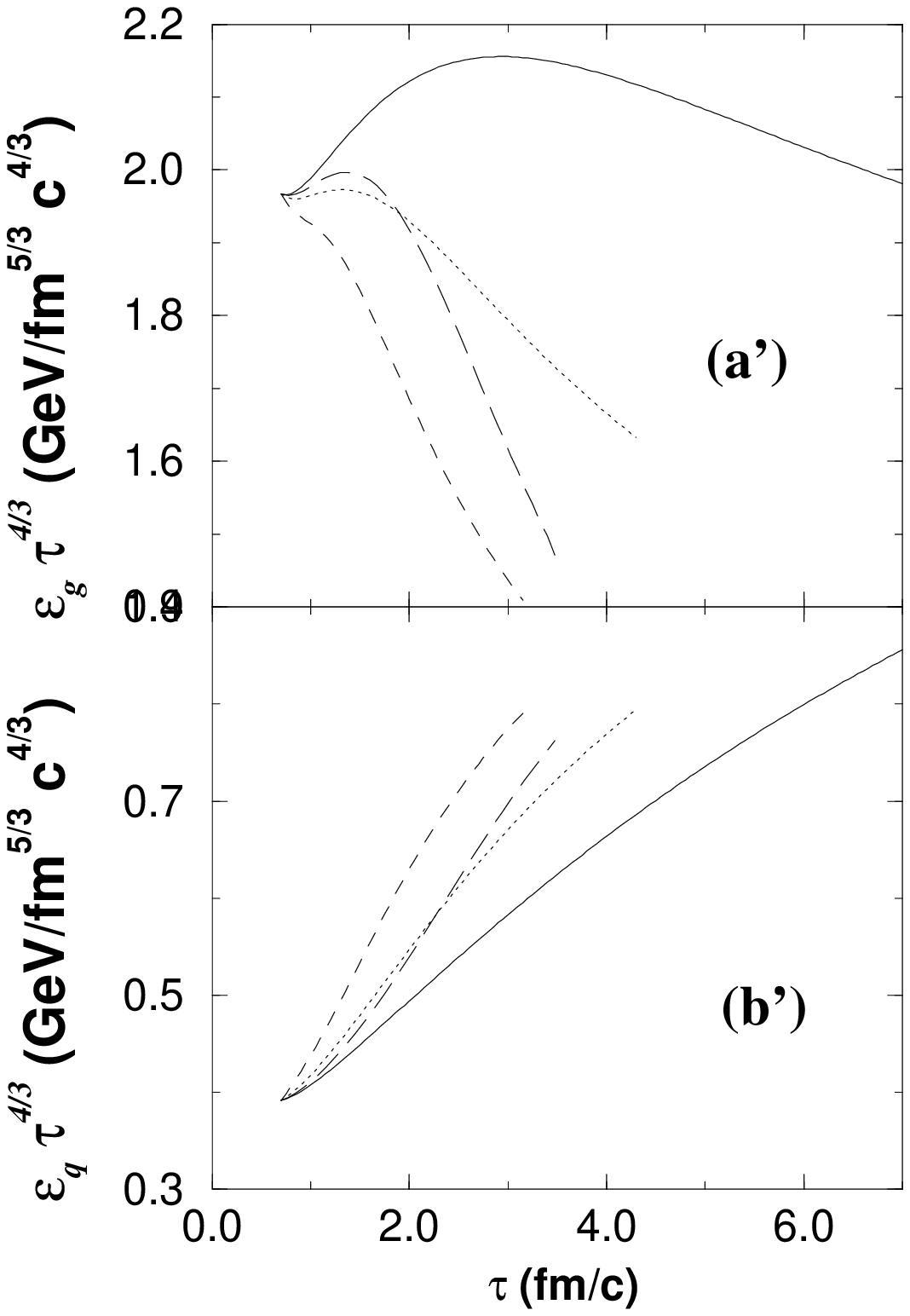,width=2.3in}}
}}
\caption{From the products $\e_i \t^{4/3}$, one can deduce
information on the effective pressure $p_{L\,i\,\eff}$
and hence the energy transfer variation with the coupling.
As before (a) and (b) are results for LHC and (a') and (b')
are those for RHIC. The couplings are assigned to the
curves in the same way as in previous figures.}
\label{f:eprod}
\efig

\bfig
\centerline{
\hbox{
{\psfig{figure=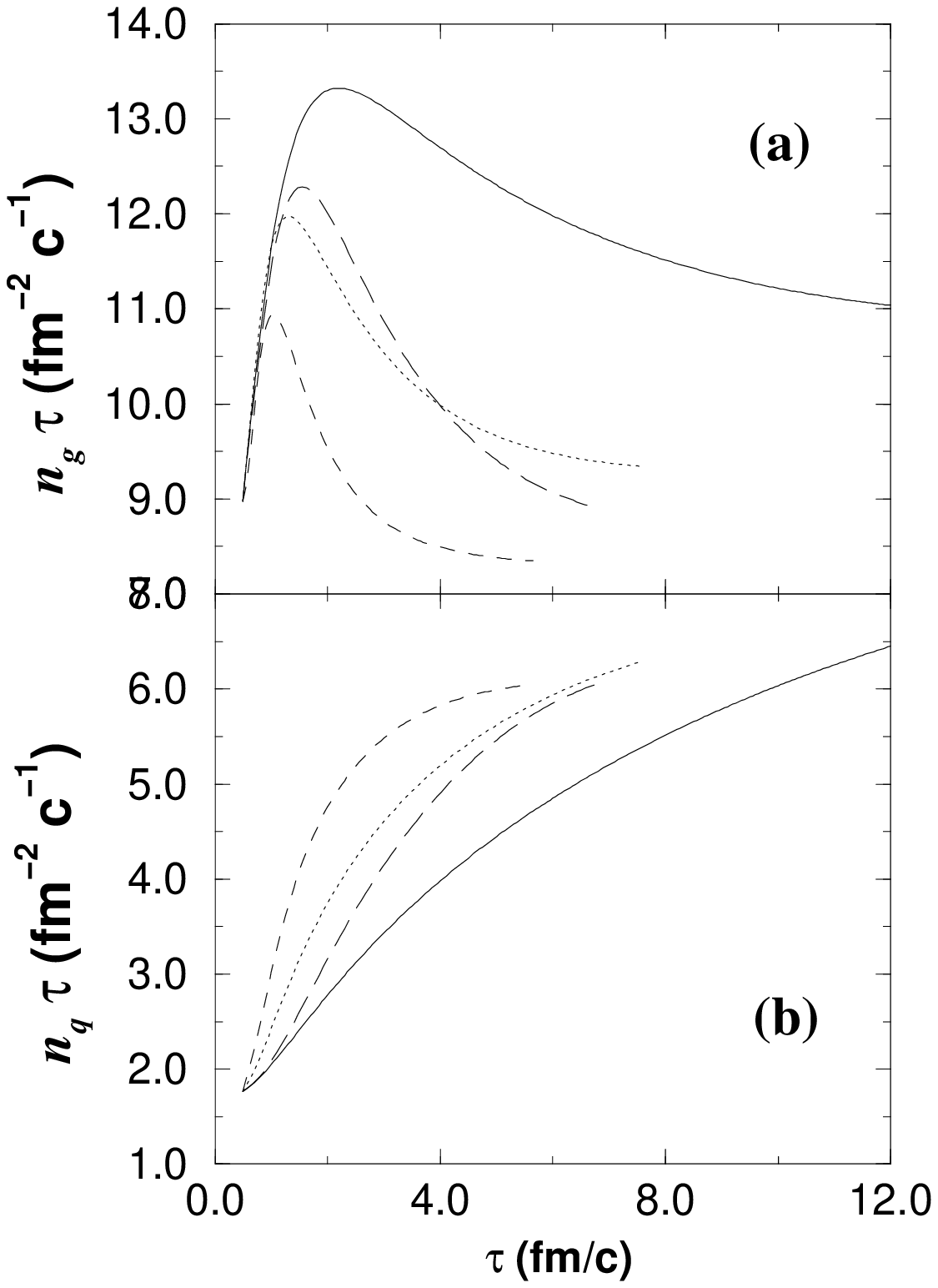,width=2.3in}} \ 
{\psfig{figure=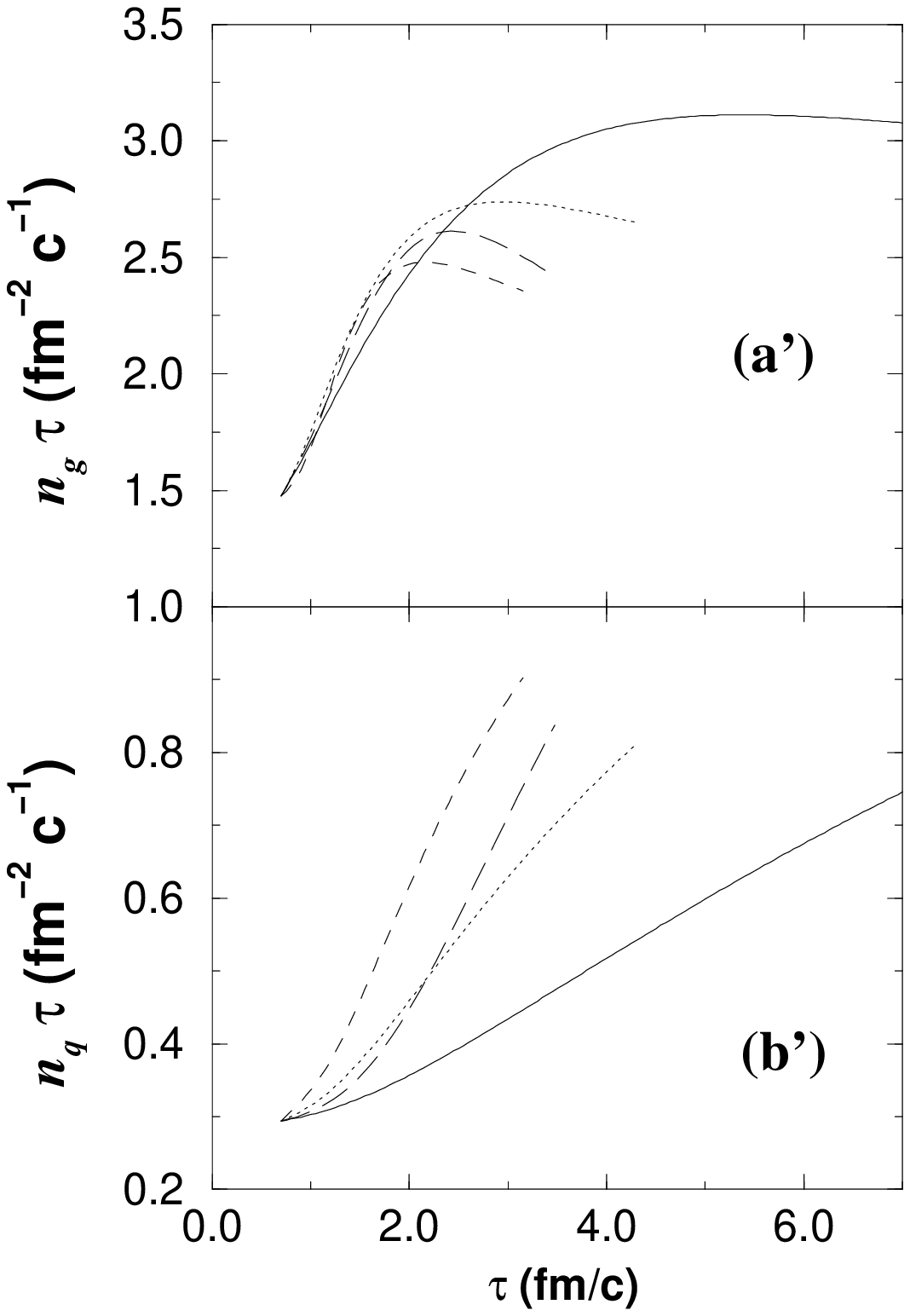,width=2.3in}}
}}
\caption{These figures show the more favorable conversion
of gluons into quark-antiquark pairs with increasing
coupling. The reduction in the produced net number of 
gluons in (a) and (a') and therefore the diminution of the 
gluon density is accompanied by the more abundance creation 
of fermion pairs shown in (b) and (b') and hence an 
increase of their density.}
\label{f:nprod}
\efig

The net creations of gluons are reduced more and more
with increasing $\a_s$ in \fref{f:nprod} (a) and (a')
by the stronger conversion process of gluons into 
quark-antiquark pairs in \fref{f:nprod} (b) and (b'). 
As seen, these result in the lowering and earlier
occuring of the peak number of gluons in figure (a) and
(a') and the drop in gluon density is accompanied
by an increase in the quark and antiquark density.
This is so because of the faster chemical equilibration 
which we have already seen with increasing coupling so 
gluons are closer to chemical equilibrium earlier which 
favors the conversion into quark-antiquark pairs. 

The last collective variable and also the most 
important one that we are interested in is the 
entropy. As we have already mentioned, in \fref{f:s}, 
one can see the product of the gluon entropy density
$s_g$ with $\t$, $s_g \t$, for various $\a_s$ 
decreases faster and just the opposite happens to the 
product $s_q \t$. They increase more rapidly with $\a_s$.
These are as expected from the results obtained so far. 
The most interesting part is however in the product of the total 
entropy density with $\t$. The more rapid equilibration 
associated with larger $\a_s$ reduces the produced entropy 
and therefore final pion multiplicity when the plasma 
eventually freezes and breaks up. This is most clear at 
LHC where the state of equilibration is much better than 
that at RHIC.  

\bfig
\centerline{
\hbox{
{\psfig{figure=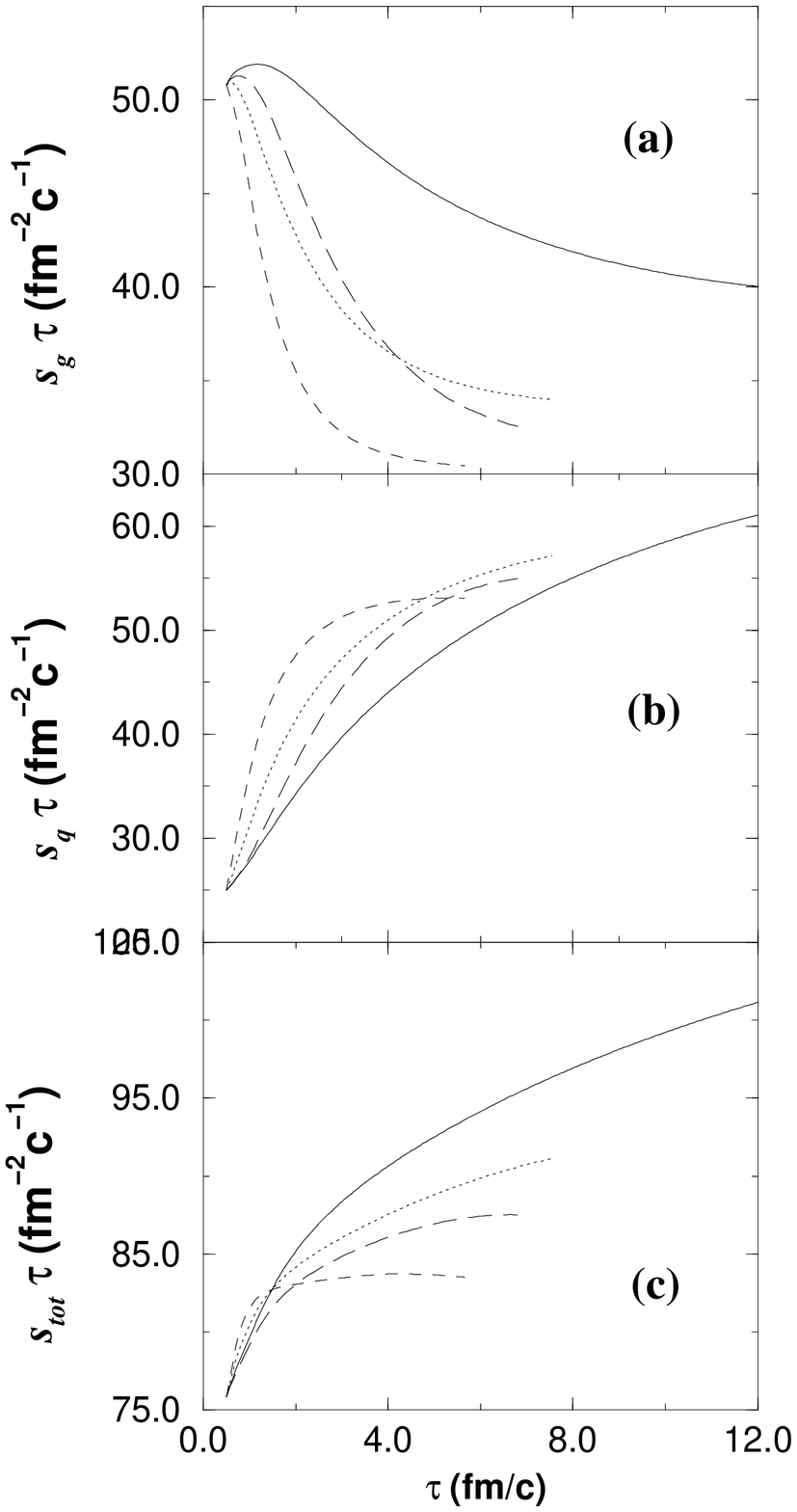,width=2.3in}} \ 
{\psfig{figure=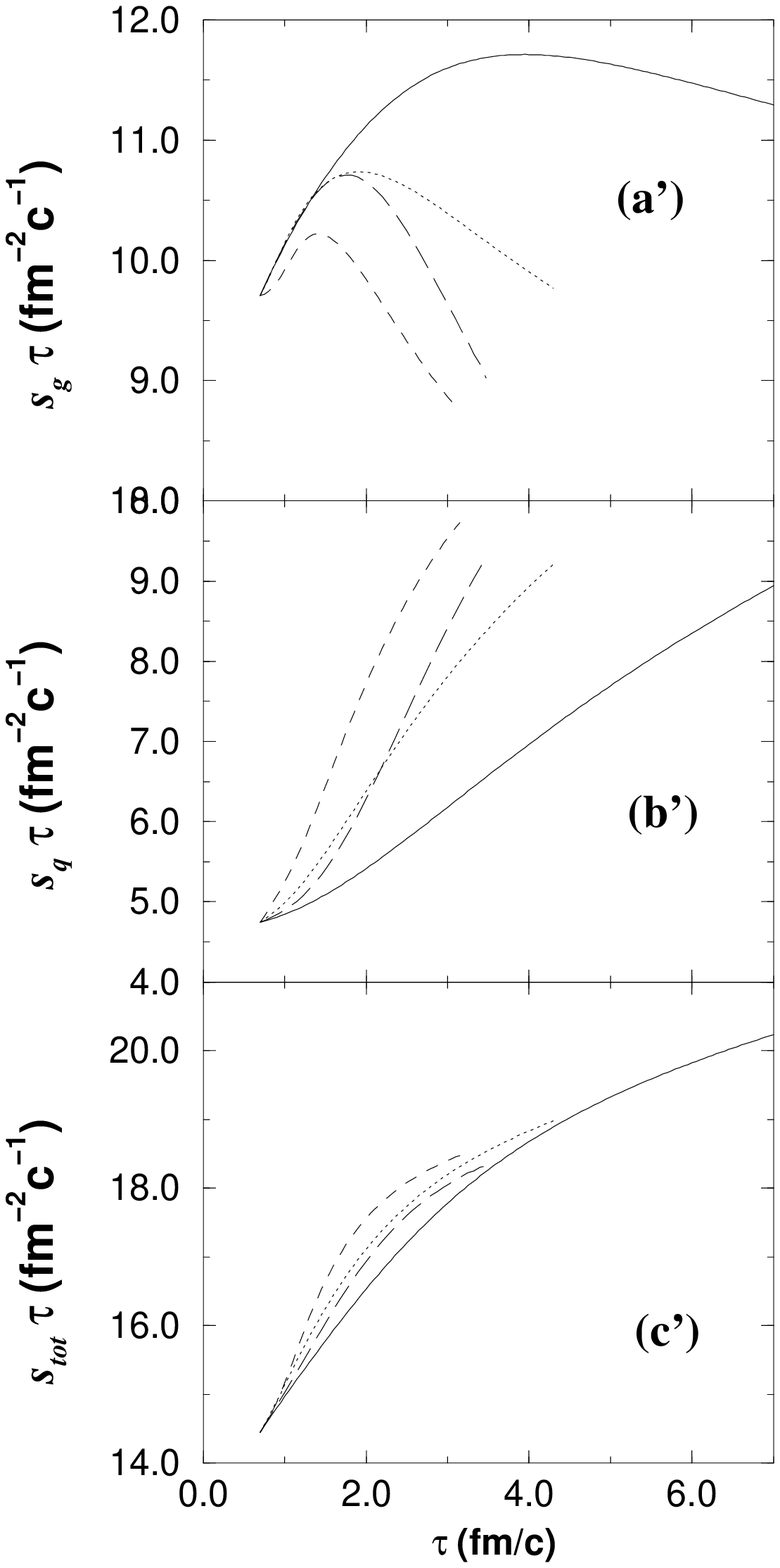,width=2.3in}}
}}
\caption{Whereas there is an increasing gain in the 
quark and antiquark entropy with increasing coupling
in (b) and (b'), the gluon entropy is reduced more
and more in (a) and (a'). The total sum is also reduced
by increasing strength of the interactions. This is much 
clearer at LHC in (c) where the state of the equilibration
is much better than that at RHIC in (c').}
\label{f:s}
\efig

From these last discussed figures, one can again see the
faster equilibration with increasing $\a_s$ already
shown in \fref{f:fug} and \fref{f:pres}. As discussed in
some details in \cite{wong2}, energy transfer and number
conversion between gluons and quarks and antiquarks
will tend to zero as equilibrium is approached,
therefore the products in \fref{f:eprod} and 
\fref{f:nprod} will tend to be independent of $\t$.
Also entropy generation will cease once equilibrium
has been attained, then the product in \fref{f:s} will
likewise progressively be independent of $\t$.
At LHC, one can see this quite clearly but 
unfortunately not so at RHIC.

In \cite{wong1,wong2}, we have discussed the 
connection of the collision time $\q$ with the stage 
of the equilibration. A large $\q$ indicates a small 
net interaction rate and {\em not} a small interaction 
rate since it is the difference between the forward 
and backward reaction which enters 
the collision terms in the Boltzmann equation. Due to 
colour, quark and antiquark interact more weakly than
gluon in general, therefore $\q_g < \q_q$. With our
initial conditions, interactions have to bring
the expanding plasma under control first before
guiding it towards equilibrium. This is manifested in
the initial rapid drop of $\q$, especially $\q_q$,
and the eventual slow rise. The initial rapid drop in
$\q$ is a response of the system to being driven out
of equilibrium by the expansion. The net interaction
rate is forced to increase rapidly until it overtakes
the expansion rate, at which point $\q$ ends its 
downward descent and begins its slow rise.
With close to equilibrium initial conditions, the 
initial drop will be absent. One can understand 
the final behavior from the calculations
of relaxation time near equilibrium 
\cite{baymetal1,baymetal2,heis}. Their known behavior near
equilibrium is $1/T$ so as the system cools, the collision
time should rise. In terms of the net interaction
rate, this rate will become slower and slower as 
equilibrium is near so $\q$ must rise. This behavior 
is correct has already been demonstrated in \cite{wong1}. 
One can see from \fref{f:coll} (a) and (b) for both $\q_g$
and $\q_q$ at LHC and similarly in the (a') and (b') 
figures at RHIC, the same patterns appear in all the 
fixed $\a_s$ results. But for the $\a_s^v$ case, something
very interesting happens. The later stage increase with $\t$
of $\q_g$ is either less fast for the gluons in (a) or 
in (a') and for the quarks $\q_q$ in (b) and (b'), 
where they even decrease further with $\t$.
This continued decrease is however different from the initial
rapid drop and it does not mean that the plasma is not
approaching equilibrium according to our reasoning
given here. As we saw in \fref{f:fug} and
\fref{f:pres}, this is not the case but rather $\a_s^v$
evolves in such a way as to compensate for the slowing 
down of the net interaction rate as equilibrium is
approached. This only happens
in a non-abelian theory with asymptotic freedom as in
the case of QCD. It does not happen if the strength of 
the interaction is fixed as in an usual ideal molecular gas 
for example. So the equilibration of the parton plasma is
helped along the way towards equilibrium by the increasing
coupling but this same phenomena will cut short the  
equilibration of the parton system as the deconfinement
phase transition begins to take place. In the case of a
first order phase transition, the equilibration will
continue in the mixed hadron-parton system and is
outside the scope of this paper.

\bfig
\centerline{
\hbox{
{\psfig{figure=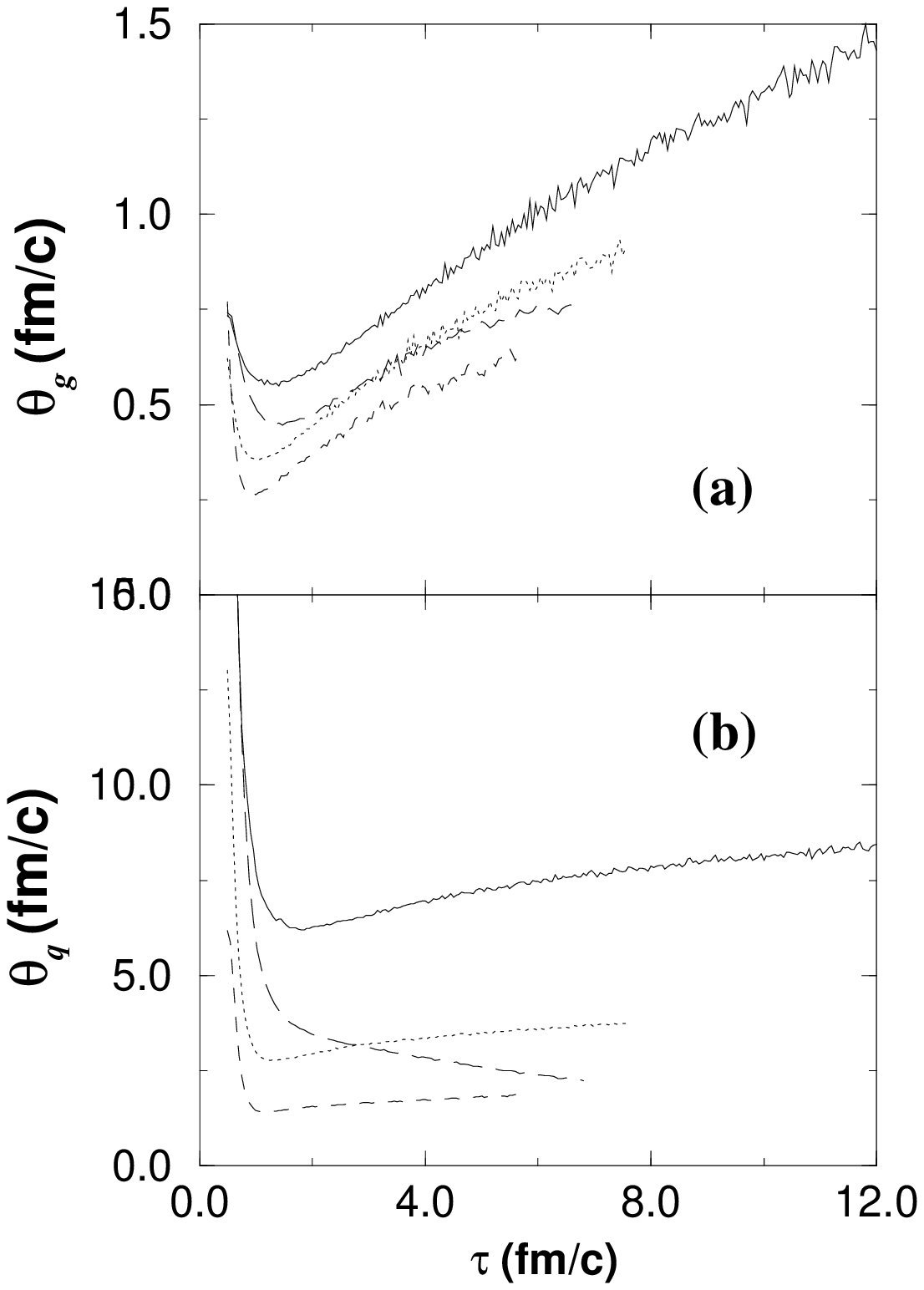,width=2.3in}} \ 
{\psfig{figure=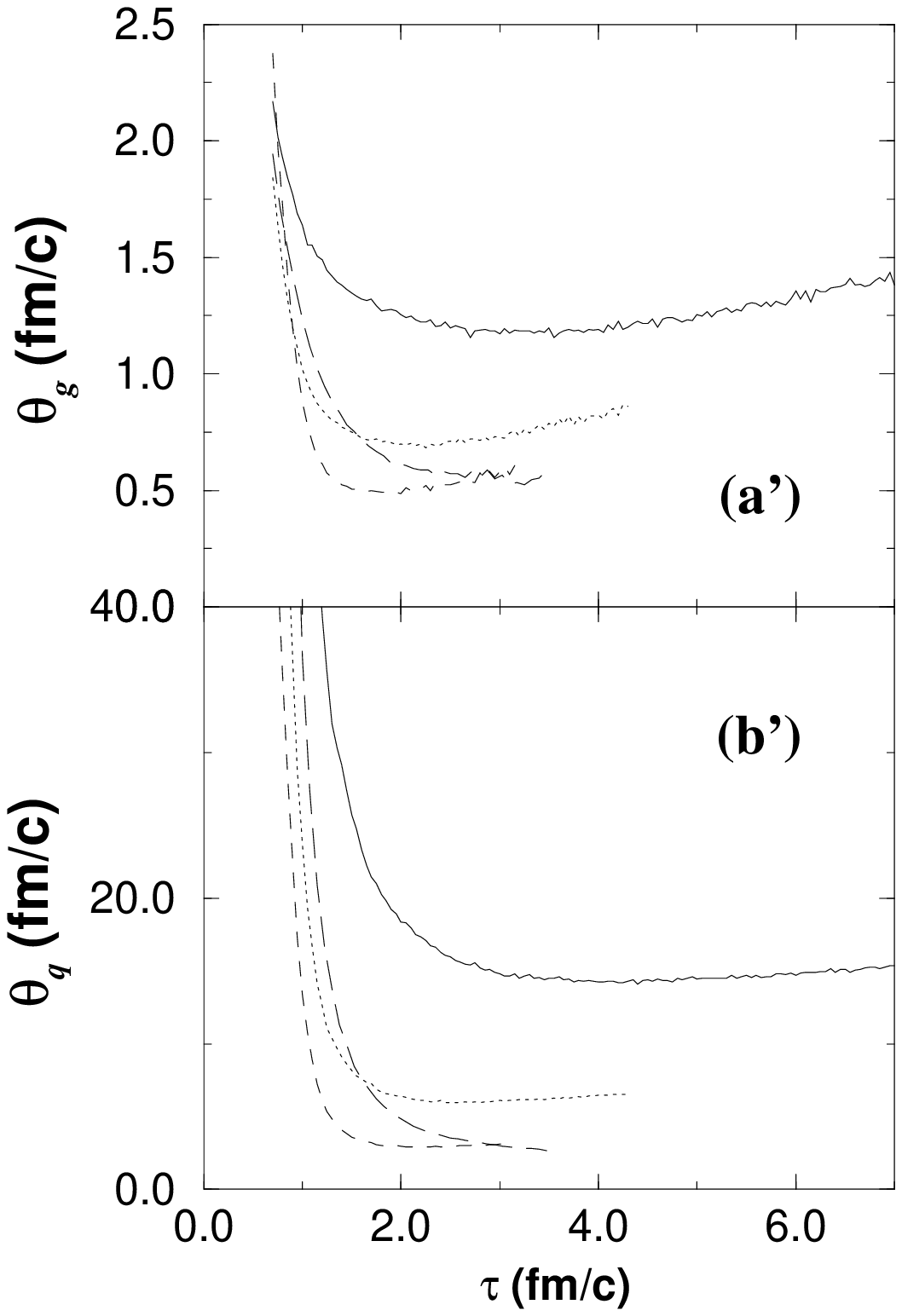,width=2.3in}}
}}
\caption{The time evolution of the collision time reflects
the state of the equilibration. The behaviors are similar
amongst the curves with constant couplings: $\a_s=0.3$ (solid),
0.5 (dotted) and 0.8 (dashed). The exceptions are the
curves with the varying coupling $\a_s^v$ (long dashed)
both at LHC (a) and (b) and at RHIC (a') and (b') which
show accelerated approach to equilibrium not found
in the equilibration of ordinary many-body system.}
\label{f:coll}
\efig

\bfig
\centerline{
\psfig{figure=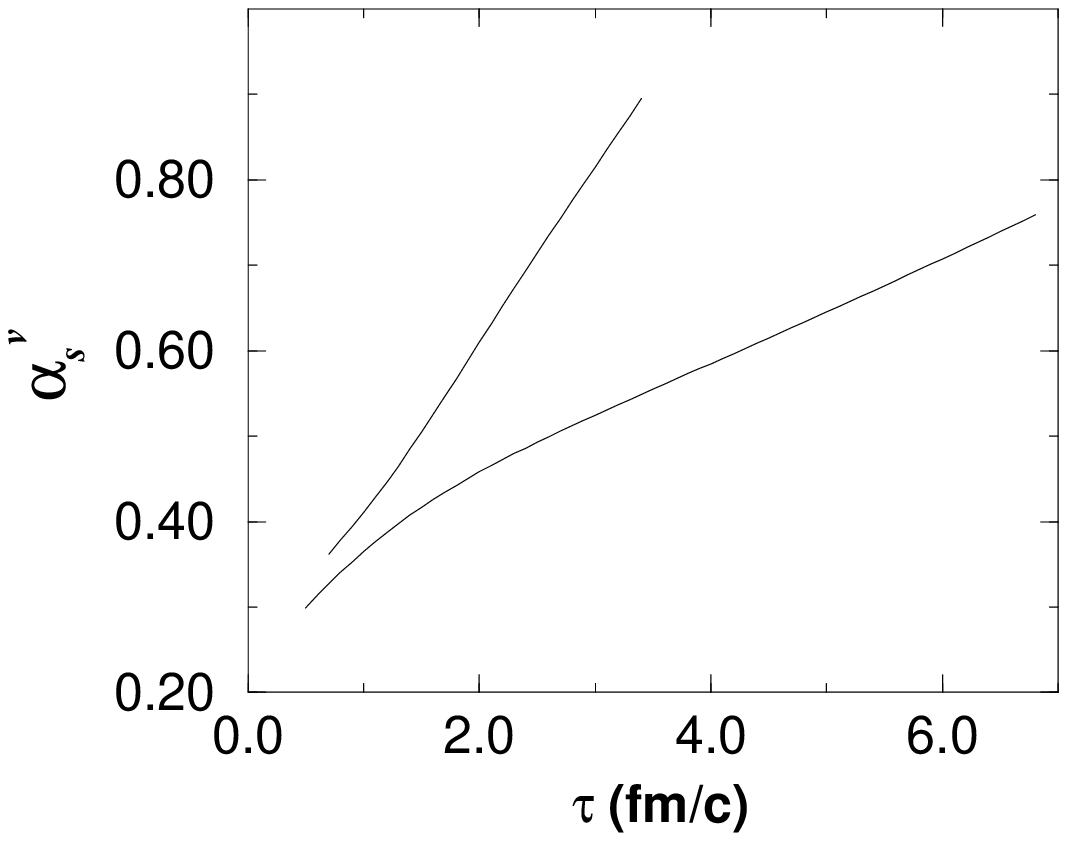,width=2.3in}
}
\caption{As the expanding parton plasma approaches 
equilibrium, the strength of the interactions also 
increases which is the basic reason for the acceleration 
in the equilibration. As seen here, perturbative 
calculations are less favorable at RHIC (top curve) 
than at LHC (bottom curve).}
\label{f:coup}
\efig

Finally, we plot the evolution of $\a_s^v$ in \fref{f:coup}.
The bottom (top) curve is for LHC (RHIC). The values of
$\a_s^v$ depend to a certain extent on the $\Lambda_{QCD}$
used. With our present choice, RHIC is seen to be less
favorable for perturbative calculations than LHC. 
Because $\a_s^v$ never exceeds 1.0, one can still use
perturbative calculation in principle. However the
values of $\a_s^v$ at later stages are uncomfortably
large. At some point already higher orders have better
be included. How these will alter our results will have
to be studied. Since initial conditions from HIJING used
here are, perhaps, some of the less favorable inputs,
one can envisage other ones that could prolong the
duration favorable for perturbative calculations.

To summarize, we have checked how equilibration is
affected by the choice of the coupling and we have
also shown that to use a fixed value of the
coupling for the whole duration of the parton
plasma right down to the deconfinement phase
transition is not consistent. The results of the
equilibration do depend on what $\a_s$ is used. In
particular, the duration of the parton phase
and the entropy are sensitive to the value of
$\a_s$. Between gluons and quarks, only the quark
final degree of equilibration is affected by the 
value of the coupling. So, in general, to let the system
decides its own strength of the interactions may be
a better choice. The best choice is then not to have
to choose at all. With a evolving coupling,
equilibration is faster and better because the now
increasingly strongly interacting parton plasma
compensates for the slowing down of the equilibration
as equilibrium is approached. This accelerated 
approach to equilibrium is not found in other many-body
system and is unique to the parton plasma in which
the interactions are described by the QCD Lagrangian.
Because the gluon end degree of equilibration does
not change much with the coupling but that of the
quark does, the second stage in the two-stage
equilibration scenario \cite{shury} will not last as long
as with a fixed coupling at $\a_s=0.3$, if the
phase transition has not already started before 
its completion.

\section*{Acknowledgements}

The author would like to thank Al Mueller for giving the
idea to this investigation and for discussion.

\vfill\break

\section*{Figure Captions}

\begin{itemize}

\item[\fref{f:ave}]{The evolution of the average parton energy 
of gluon (solid line) and quark or antiquark (dashed line) at
(a) LHC and (b) RHIC with $\a_s=0.3$.}

\item[\fref{f:fug}]{Chemical equilibration of (a) gluons and 
(b) quarks with various values for the coupling: $\a_s=0.3$ 
(solid), 0.5 (dotted), 0.8 (dashed) and $\a_s^v$ (long dashed) 
at LHC. The (a') and (b') figures are the same at RHIC.
Increasing coupling improves the quark final degree of chemical
equilibration much more than that of the gluon.}

\item[\fref{f:pres}]{Using the ratios of the longitudinal 
pressure and a third of the energy density to the transverse 
pressure to check for isotropy in momentum distribution and 
therefore kinetic equilibration. The bottom (top) set of four 
curves in each figure is for the pressure (energy density) to
pressure ratio. The assignments of the coupling to the
curves are $\a_s=0.3$ (solid), 0.5 (dotted), 0.8 (dashed)
and $\a_s^v$ (long dotted). Figures (a) and (a') are
for gluons and (b) and (b') for quarks at LHC and at RHIC
respectively. Faster kinetic equilibration is seen
everywhere with larger $\a_s$ but improvement in the 
final degree of thermalization is essentially reserved 
for the fermions and not for the gluons.}

\item[\fref{f:temp}]{The time variations of the estimated 
temperatures of (a) and (a') gluons and (b) and (b') quarks 
at LHC and at RHIC respectively. These temperatures drop faster
with increasing coupling. The different values of the
coupling are assigned to the curves in the same
way as in \fref{f:fug} and \fref{f:pres}.}

\item[\fref{f:eprod}]{From the products $\e_i \t^{4/3}$, one can 
deduce information on the effective pressure $p_{L\,i\,\eff}$
and hence the energy transfer variation with the coupling.
As before (a) and (b) are results for LHC and (a') and (b')
are those for RHIC. The couplings are assigned to the
curves in the same way as in previous figures.}

\item[\fref{f:nprod}]{These figures show the more favorable 
conversion of gluons into quark-antiquark pairs with increasing
coupling. The reduction in the produced net number of 
gluons in (a) and (a') and therefore the diminution of the 
gluon density is accompanied by the more abundance creation 
of fermion pairs shown in (b) and (b') and hence an 
increase of their density.}

\item[\fref{f:s}]{Whereas there is an increasing gain in the 
quark and antiquark entropy with increasing coupling
in (b) and (b'), the gluon entropy is reduced more
and more in (a) and (a'). The total sum is also reduced
by increasing strength of the interactions. This is much 
clearer at LHC in (c) where the state of the equilibration
is much better than that at RHIC in (c').}

\item[\fref{f:coll}]{The time evolution of the collision time 
reflects the state of the equilibration. The behaviors are 
similar amongst the curves with constant couplings: $\a_s=0.3$ 
(solid), 0.5 (dotted) and 0.8 (dashed). The exceptions are 
the curves with the varying coupling $\a_s^v$ (long dashed)
both at LHC (a) and (b) and at RHIC (a') and (b') which
show accelerated approach to equilibrium not found
in the equilibration of ordinary many-body system.}

\item[\fref{f:coup}]{As the expanding parton plasma approaches 
equilibrium, the strength of the interactions also 
increases which is the basic reason for the acceleration 
in the equilibration. As seen here, perturbative 
calculations are less favorable at RHIC (top curve) 
than at LHC (bottom curve).}

\end{itemize}

\end{document}